# Electronic structure of $Li_2Pd_3B$ and $Li_2Pt_3B$

Sharat Chandra[§], S. Mathi Jaya and M.C. Valsakumar

*Materials Science Division, Indira Gandhi Centre for Atomic Research, Kalpakkam, 603102,*

*Tamil Nadu, India*

**Abstract**

$Li_2Pd_3B$ is known to be superconducting, while the isotypical $Li_2Pt_3B$ compound is not. Electronic structures of $Li_2Pd_3B$ and $Li_2Pt_3B$ have been calculated in order to obtain an insight into this surprising difference, through an analysis of the differences in the band structures. The electronic structures of these systems were obtained using the Full Potential Linear Augmented Plane Wave plus local orbitals (FP-LAPW+lo) method and it was found that four bands cross the Fermi level ($E_F$). Out of these four bands, only two bands contribute significantly to the density of states at the $E_F$. One of these bands is a hole band and the other an electron band. Thus at least a two-band model is required for studying the electronic properties of the Pd and Pt compounds. These two bands are rather narrow and hence the coulombic correlations effects can be significant.

---

[§] Corresponding author email: sharat@igcar.ernet.in

I. **Introduction**

The discovery of superconductivity in $MgB_2$ [1] and subsequently in $MgCNi_3$ [2] and $Li_2Pd_3B$ [3,4] has evoked substantial experimental and theoretical studies on various properties of these compounds. The studies on $MgB_2$ have revealed that the electron-phonon coupling strength in this compound is rather strong and it is believed that this compound is a phonon-mediated superconductor. However the large superconducting transition temperature ($T_C$) of this compound remains as a puzzle. The studies on $MgCNi_3$ revealed that it is a two-band superconductor [5] and the presence of the van Hove singularity near the Fermi energy in the density of states is believed to have a strong role in the promotion of superconductivity in this system [5,6]. The superconductivity in $Li_2Pd_3B$ is proposed to be due to the Pd-d electrons [7] and these d-electrons are expected to have strong coulombic correlation effects that are likely to be dominant in the electronic properties of this system. The isotypical compound $Li_2Pt_3B$, on the other hand, is found to be nonsuperconducting down to 4 K. This surprising difference in the properties of these materials has lead us to undertake calculation of the electronic structures of these systems which can provide information concerning the width and character of the bands that cross the Fermi level, the density of states at the $E_F$ and its contribution from different orbitals. We have evaluated the electronic structure in order to quantify the differences in its electronic structure of $Li_2Pd_3B$ and $Li_2Pt_3B$ and analyzed the possible causes of the absence of superconductivity in $Li_2Pt_3B$. We describe our results in the following sections along with a brief description of the details of calculations.

**II. Details of calculation**

The crystal structure of the $Li_2Pd_3B$ has been plotted using the VENUS package [8]. Detailed electronic structure of $Li_2Pd_3B$ and $Li_2Pt_3B$ compounds were calculated using the Full Potential Linear Augmented Plane Waves plus local orbitals (FP-LAPW+lo) code WIEN2k [9] using the Generalized Gradient Approximation (GGA) of the exchange correlation functional due to Perdew, Burke and Ernzerhof (PBE96) [10]. The self-consistent field (SCF) cycles were performed using 1000 k-points in the irreducible wedge of the Brillouin Zone, till convergence was achieved in energy, force and charge. After the completion of SCF cycles the energy convergence and charge separation distance were less than $10^{-6}$ Rydberg (Ry) and $10^{-6}$ atomic units respectively and the calculated forces between the atoms were lower than 5 mRy/a.u. The Fermi surfaces have been calculated and plotted using the X-window crystalline structures and densities program XCrySDen [11]. Experimental lattice constants and atom positions [12] were used for all the calculations and it seems that the experimental crystal structure is very nearly optimized, as the forces on the atoms are < 5 mRy/a.u. after completion of the SCF cycle.

### III. Crystal structure

The crystal structure of $Li_2Pd_3B$ is shown in Fig.1. Both $Li_2Pd_3B$ and $Li_2Pt_3B$ are isotypical and crystallize in the **P**$4_3$32 structure (Space Group No. 212), with 4 formula units per unit cell [12]. The lattice constant of $Li_2Pd_3B$ is 0.67534(3) nm and that of $Li_2Pt_3B$ is 0.67552(5) nm. For the $Li_2Pd_3B$ structure, Li is at (x,x,x) position with x=0.3072 (Wyckoff multiplicity 8c), Pd is at (1/8,y,1/4-y) position with y=0.30417 (Wyckoff multiplicity 12d) and B is at special position (7/8,3/8,1/8), (Wyckoff multiplicity 4b). The corresponding fractional atom coordinates in the isotypical $Li_2Pt_3B$ structure are for Li, x=0.2930 and for Pt, y=0.3079. Fig.1(a) represents the Pd-B coordination as an distorted octahedron. This coordination octahedron has a volume of $11.988 \times 10^{-3}$ nm$^3$. Baur [13] and Robinson *et.al.*, [14] have defined three distortion indices for any polyhedron with a central atom. These are: the distortion index, $D = \frac{1}{n}\sum_{i=1}^{n}\frac{|l_i - l_{av}|}{l_{av}}$, the quadratic elongation, $<\lambda> = \frac{1}{n}\sum_{i=1}^{n}\frac{l_i}{l_o}$ and the bond angle variance, $\sigma^2 = \frac{1}{m-1}\sum_{i=1}^{m}(\varphi_i - \phi_o)^2$, for the distorted polyhedron. The coordination number of the central atom is *n* (=6 for B). $l_i$ is the distance from the $i^{th}$ coordinating atom to the central atom, i.e., the $i^{th}$ bond length, $l_{av}$ the average bond length, $l_o$ the bond length in a regular polyhedron of same volume. *m* is given by the number of bond angles, or, (3/2)×(number of faces in polyhedron). $\varphi_i$ is the $i^{th}$ bond angle and $\varphi_o$ the ideal bond angle for the regular polyhedron (90º for an regular octahedron). As all the bond lengths between the central B atom and the Pd atoms at vertices are 0.21315 nm, $D=0$, and $<\lambda>$ and $\sigma^2$ can be calculated to be 1.0508 and 169.168 degree$^2$ respectively for the structure shown in Fig.1(a). Thus it is very apparent that the extent of distortion in the polyhedron is not so much in the bond distances as in the bond angles. This leads us to view the structure in a different way. The alternate view of the $Li_2Pd_3B$ structure is shown in Fig.1(b). This is just the unit cell shown in Fig.1(a) viewed in the (111) orientation. Immediately it is apparent that the distorted octahedron breaks up into two tetrahedra of equal volume of $2.188 \times 10^{-3}$ nm$^3$ that are formed by the Pd and B atoms and are rotated with respect to each other by 35.68º. Note that if the angle of rotation was 60º, then a regular octahedron would have been obtained. The central B atom is shared between two tetrahedra which are denoted as Pd and Pd' for our convenience. The B-Pd or B-Pd' bond distance is 0.2131 nm, that between Pd-Pd or Pd'-Pd' is 0.2788 nm while the Pd-Pd' distance is 0.2964 nm. The angles between the base Pd atoms are 60º, while the corresponding angles between the B atom at the vertex and the Pd atoms in the base (Pd-B-Pd) are equal to 81.692º.

This analysis of the structure leads us to a newer interpretation of the bonding in these materials. As the Pd-Pd' distance is 17% larger than the Pd-Pd distance, only the Pd-Pd distance is comparable to the nearest neighbor bond distance in metallic palladium (0.275 nm). As there is strong covalent bonding between the B and Pd atoms, it is expected that only the Pd-B and Pd-Pd bonds and not the Pd-Pd' bonds should play a significant role in determining the electronic properties of these materials. As the Li-Li distance is 0.2557 nm, which is much less than that in Li metal (0.304 nm), there is a complete transfer of electrons to the Pd-B sublattice. Hence, Li atoms serve only to stabilize the structure after they have donated the electrons.

The above discussion also holds true for the $Li_2Pt_3B$ compound, as the variations in the interatomic distances and bond angles are similar to the $Li_2Pd_3B$ compound. The $Pt_6B$ octahedral volume is $12.088 \times 10^{-3}$ nm$^3$ and $D$, $<\lambda>$ and $\sigma^2$ can be calculated to be zero, 1.0543 and 184.218 degree$^2$ respectively. The B-Pt, Pt-Pt and Pt-Pt' bond distances are 0.2141 nm, 0.2756 nm and 0.3026 nm respectively. The corresponding tetrahedral volume is $2.246 \times 10^{-3}$

nm$^3$. The angle of rotation between the two tetrahedra is 36.79º and the Pt-B-Pt angle is 80.444º.

## IV. Results and Discussion

### 1. Li$_2$Pd$_3$B

The calculated band structure of Li$_2$Pd$_3$B is shown in Fig.2(a). The bands that cross the Fermi level are shown in thicker lines. It is seen from the figure that four bands cross the Fermi level. In order to evaluate the contribution to the density of states at the Fermi energy (N(E$_F$)) from these bands, the band by band density of states of these four bands were calculated and they are shown in Fig.3(a). We can see that the contribution to the N(E$_F$) is predominantly due to two bands (the second and third bands). The other two bands (the first and fourth bands) give negligibly small contribution to the N(E$_F$). Thus the two bands which contribute significantly to the N(E$_F$) are sufficient in describing the properties of this system. Band 2 is almost occupied, hence is a hole band while band 3 is almost empty and so is an electron band. The widths of these two bands are 490 meV and 599 meV respectively. In comparison, the widths of the band that cross the Fermi energy in MgCNi$_3$ [5,15] has been shown to be ~0.85 and 1.75 eV. As the two bands are quite narrow in Li$_2$Pd$_3$B, the coulombic correlation effects in these bands can be significant. It may be noticed that a van Hove singularity very close to the E$_F$ is present in the electron band. It is just 4 meV above the Fermi level. The density of states (DOS) and the contribution from different atoms are shown in Fig.4(a). It is seen from the figure that the density of states around E$_F$ is predominantly due to the Pd atoms and major contribution comes from the d-orbitals. Pd atoms contribute 3.755 states/eV/cell, B atoms 1.077 states/eV/cell and Li atoms contribute 0.382 states/eV/cell at the Fermi level. The s, p and d-orbitals of the Pd atoms contribute 0.2208, 0.3744 and 3.1404 states/eV/cell, while the p-orbitals of the B atom contribute 0.9860 states/eV/cell. Maximum contribution to the Pd d-states at the Fermi level comes from the $d_{x^2-y^2}$ orbitals.

It has been discussed in literature that the important feature of the crystal structure of this compound is the distorted $Pd_6B$ octahedral network [7]. This distortion is thought to create a difference in the charge density among the Pd atoms. In order to verify whether such a difference in the charge density of the Pd atoms exists or not, we have plotted in Fig.5(a) the charge density of this compound in the (311) plane. This plane passes through the Pd, Pd', B and Li atoms marked in Fig.5(a). It can be seen from the charge density plots that the bonding between the Pd and B atoms is covalent in nature, whereas that between the Pd atoms, which are separated by 0.2788 nm is metallic. The nature of bonding between the Li and Pd atoms is ionic like as can be seen from the closed nature of the contours. It is also clearly seen that the charge density on both the Pd and Pd' atoms are identical. This behavior also follows from the discussion on the geometry of the crystal structure presented above.

## 2. $Li_2Pt_3B$

The calculated band structure of $Li_2Pt_3B$ is shown in Fig.2(b). The bands crossing the Fermi level are shown as thick lines. Here again four bands cross the Fermi level and the contribution to the $N(E_F)$ is predominantly due to the second and third band. The partial density of states of the bands crossing the Fermi level is shown in Fig.3(b). The widths of the second and third bands are respectively 381 meV and 531 meV. As the widths of these bands are small the coulombic correlation effects can be significant in this system too. A van Hove singularity is present in the hole band at 52 meV, below the $E_F$. The total density of states for $Li_2Pt_3B$ and the contribution to total DOS from different atoms is shown in Fig.4(b). The overall magnitude of the total DOS is observed to be lower in $Li_2Pt_3B$ than that in $Li_2Pd_3B$, which points to the sharper nature of the Pd 4d-bands in the $Li_2Pd_3B$ compound. Also, we observe from the band structure that there is a gap opening up at ~1eV energy. This gap is absent in the $Li_2Pd_3B$ band structure. The contributions to the total DOS at the Fermi energy are: 4.466 states/eV/cell by Pt, 1.203 states/eV/cell by B and 0.243 states/eV/cell by Li atoms. 0.2256, 0.3132 and 3.8604 states/eV/cell are contributed by the s, p and d-orbitals of the Pt atoms, while the p-orbitals of the B atoms contribute 1.0732 states/eV/cell. Maximum contribution to the Pt d-states at Fermi level is from $d_{x^2-y^2}$ states. The electron density distribution in the (311) plane for $Li_2Pt_3B$ is shown in Fig.5(b). The atoms have been labeled as in Fig.5(a). It is seen that there is similarity in the electron density distribution in the $Li_2Pd_3B$ and $Li_2Pt_3B$ compounds to a large extent. The bonding between the Pt and B atoms is covalent, that between the Pt atoms is metallic and that between Li and Pt atoms is ionic in nature.

As can be seen from Fig.4, the density of states at Fermi level is larger in $Li_2Pt_3B$ than that in $Li_2Pd_3B$. In the BCS scenario, such an increase in $N(E_F)$ should lead to an enhancement of the $T_C$. However, there could be a reduction in $T_C$ in view of the fact that Pt atom is heavier than the Pd atom. An estimate of the lower limit to $T_C$ of $Li_2Pt_3B$ can be obtained by ignoring the enhancement that may arise from the increase in $N(E_F)$ and ascribing the entire isotope effect to the Pd/Pt atoms alone [16]. Then we get, $T_C(Li_2Pt_3B) \geq T_C(Li_2Pd_3B) \cdot \sqrt{M_{Pd}/M_{Pt}} \geq 5.9K$, where $M_{Pd}$ and $M_{Pt}$ are the atomic masses of Pd and Pt atoms. Here we have also ignored the modification to the isotope effect due to Coulomb interactions, which results in a further reduction of the isotope effect [17]. In view of this argument, the nonoccurrence of superconductivity down to 4 K in $Li_2Pt_3B$ would imply that this compound is not a typical BCS superconductor.

## V. Fermi Surfaces and van Hove singularity

The sheets of the Fermi surface (in the first Brillouin zone) originating from band 2 of $Li_2Pd_3B$ and $Li_2Pt_3B$ are shown in Fig.6(a) and 6(b) respectively. It can be seen that this sheet of Fermi surface has a quite complex shape and the central segments are similar to a large extent in both the $Li_2Pd_3B$ and $Li_2Pt_3B$ compounds. Disconnected features are seen in the small pockets near the R point of the Brillouin zone for both the $Li_2Pd_3B$ and $Li_2Pt_3B$ and near X point in $Li_2Pt_3B$. All the states enclosed by the complex surface segments are hole states in both the cases. Fermi surfaces for the band 3 are shown in Fig.7(a) and 7(b) respectively. The Fermi surface in case of $Li_2Pd_3B$ (Fig.7(a)) exhibits disconnected pockets along the X and M points and along M point alone in the case of $Li_2Pt_3B$ (Fig.7(b)). We observe drastic differences in the segments that surround the Γ point. In the case of $Li_2Pd_3B$, the segment is quite stretched with a six-leaf structure that has no states at the Γ point, whereas in $Li_2Pt_3B$, it encloses the Γ point and is centered on it in a small region.

The constant energy surfaces corresponding to the energy of the van Hove singularity are plotted in Fig.8(a) for $Li_2Pd_3B$ (4 meV above the $E_F$) and in Fig.8(b) for $Li_2Pt_3B$ (52 meV below the $E_F$). It can be seen that there are significant differences from the Fermi surfaces plotted in figs.6 and 7. In $Li_2Pd_3B$, the van Hove singularity arises from the structure in the vicinity of Γ point in the region where the six-leaf surface segments join together. This singularity is two dimensional in nature, as it arises from a surface that is saddle like in shape and is associated with a crossover between the electron-like and hole-like surface sections [6]. Because this shape is quite flat in the Γ-X direction and concave in nature in Γ-M direction, nesting of the surface is not significant in the case of $Li_2Pd_3B$.

It is well known that the proximity of the van Hove singularity to the Fermi level can cause Fermi surface instabilities such as superconductivity, spin density wave and charge density wave [18,19]. The relevance of van Hove singularity to the high temperature superconductivity has been well documented in a review by R.S. Markiewicz [18]. It has also been discussed in the case of $MgC_xNi_3$ ($T_C$ = ~8 K) [5,6,15]. Optimal $T_C$ is obtained for the C site (x) occupancy of 0.96 [15,20]. One of the curious facts about $MgC_xNi_3$ is the destruction of $T_C$ when x<0.9 [20]. A supercell calculation with x=0.875 showed a drastic reduction in the strength of the van Hove singularity peak without any significant change in $N(E_F)$ [5]. This argument bolsters the belief that the presence of van Hove singularity facilitates superconductivity in $MgC_xNi_3$. In the light of above arguments, it can be expected that the presence of van Hove singularity just 4 meV above the $E_F$ in $Li_2Pd_3B$ plays an important role in promotion of superconductivity. Due to the saddle nature of the constant energy surface in the region of van Hove singularity, high mass Pd 4d electrons are expected to form the superconducting pairs and Pd-Pd hopping can be important for their transport. Because the van Hove singularity is situated 4 meV above the Fermi level, electron doping might move the Fermi level to the energy position of the singularity peak. This can be an important way to study the nature of the conduction and to induce 2D instabilities in this system.

The nature of the constant energy surface of $Li_2Pt_3B$ in Fig.7(b) is very different from that of $Li_2Pd_3B$. No signature of the singularity is seen in the band structure of $Li_2Pt_3B$. This implies that the singularity does not arise due to the contributions from the surface in high symmetry directions plotted in the band structure. Due to the complex nature of the constant energy surface, it is not readily apparent as to which portion of the surface gives rise to the van Hove singularity. But we observe that large potions of the surface are extremely flat in the regions that form the edges and the mouth of the central structure in Fig.7(b). These symmetrical segments that lie in between the $\Gamma$-X and $\Gamma$-M directions give rise to extensive nesting as large portions of the surface can be translated into each other without any rotation. Even though the presence of the van Hove singularity should help in observation of $T_C$ in the Pt compound as in the case of Pd compound, extensive nesting can give rise to band antiferromagnetic instabilities [18] and can thus overwhelm the incipient superconductivity in this system. This type of nesting results in a sharp peak in the joint density of states of the system and its signature can be seen in spin (or charge) susceptibility [18]. The van Hove singularity in $Li_2Pt_3B$ is situated at 52 meV below the $E_F$ in the hole band. Even though this is comparable to the position of van Hove singularity in $MgCNi_3$, it is still one order of magnitude more than that in the Pd compound. An experimental technique such as nuclear magnetic relaxation can be used profitably to characterize these materials. Both static and dynamic measurements of the magnetic susceptibility can throw much light on the actual nature of the interactions taking place in both the $Li_2Pd_3B$ and $Li_2Pt_3B$ compounds.

## VI. Conclusions

The electronic structure of isotypical compounds $Li_2Pd_3B$ and $Li_2Pt_3B$ have been calculated using the FP-LAPW+lo method. A new interpretation of the structure of the two compounds has been given. It is shown that the distorted $Pd_6B$ octahedron can be decomposed into two regular tetrahedra rendering all the Pd atoms equivalent. This is also borne out by the calculated electron density distribution where all the Pd atoms are shown to have the same electron density. Even though four bands cross the Fermi level in both the cases, only two bands contribute significantly to the density of states at the $E_F$. Hence two-band model is sufficient to account for the properties of these compounds. The Fermi surfaces of these two bands are quite complex. van Hove singularity peaks are observed to be present near the $E_F$ in the density of states of the two compounds, 4 meV above the $E_F$ in $Li_2Pd_3B$ and 52 meV below the $E_F$ in $Li_2Pt_3B$. Even though van Hove singularity is present in both the cases, the nature of the interactions taking place in both the compounds is quite different, as there is extensive nesting of the constant energy surface in the case of $Li_2Pt_3B$. This can have a bearing on the observation of superconductivity in these compounds.

**Figure Captions**

Fig.1: The crystal structure of $Li_2Pd_3B$ viewed in the (100) orientation (a) and in the (111) orientation (b). The distorted octahedron centered on the B atom shown in (a) can be viewed as comprising of two tetrahedra with the B atom shared between two neighboring tetrahedra and Pd atoms at the other three vertices of the tetrahedra. Even though the Pd atoms that form the two tetrahedra are equivalent, they have been marked as Pd and Pd' in (b) just to differentiate them.

Fig.2: The band structure of $Li_2Pd_3B$ (a) and $Li_2Pt_3B$ (b). Four bands are seen to cross the Fermi level in both $Li_2Pd_3B$ and $Li_2Pt_3B$.

Fig.3: The partial density of states of the four bands crossing the Fermi level shown with thick lines in the band structure of $Li_2Pd_3B$ (a) and $Li_2Pt_3B$ (b). Only two bands out of the four are seen to contribute significantly to the partial DOS at the Fermi level.

Fig.4: The total density of states of (a) $Li_2Pd_3B$ and (b) $Li_2Pt_3B$. Maximum contribution to the DOS comes from d orbitals of the Pd and Pt atoms. While B atoms contribute a small percentage, Li atom contribution is almost negligible. The maximum contribution to the total DOS at the Fermi level is from $d_{x^2-y^2}$ orbitals.

Fig.5: Electron density distribution in (a): $Li_2Pd_3B$ and (b): $Li_2Pt_3B$ compounds in the (311) plane. The Pd and Pt atoms are marked as Pd, Pd', Pt and Pt' following the convention of Fig.1. Li and B atoms that lie in the planes are also marked in the figure. Other Pd, Pt and Li atoms that are at some distance from the plane are also seen in the figure. It is seen that the magnitude of the electron density at the Pd, Pd' and Pt, Pt' atom sites is equal. Also, the electron density distribution is similar in both cases.

Fig.6: The Fermi surface corresponding to the second band that crosses the Fermi level in (a) $Li_2Pd_3B$ and (b) $Li_2Pt_3B$ compounds. Fermi surfaces have been plotted in the first Brillouin zone.

Fig.7: The Fermi surface corresponding to the third band that crosses the Fermi level in (a) $Li_2Pd_3B$ and (b) $Li_2Pt_3B$ compounds.

Fig.8: Topology of the constant energy surface at the energy of the van Hove singularity. (a) In $Li_2Pd_3B$: 4 meV above the Fermi energy in band 3 and (b) In $Li_2Pt_3B$: 52 meV below the Fermi energy in band 2.

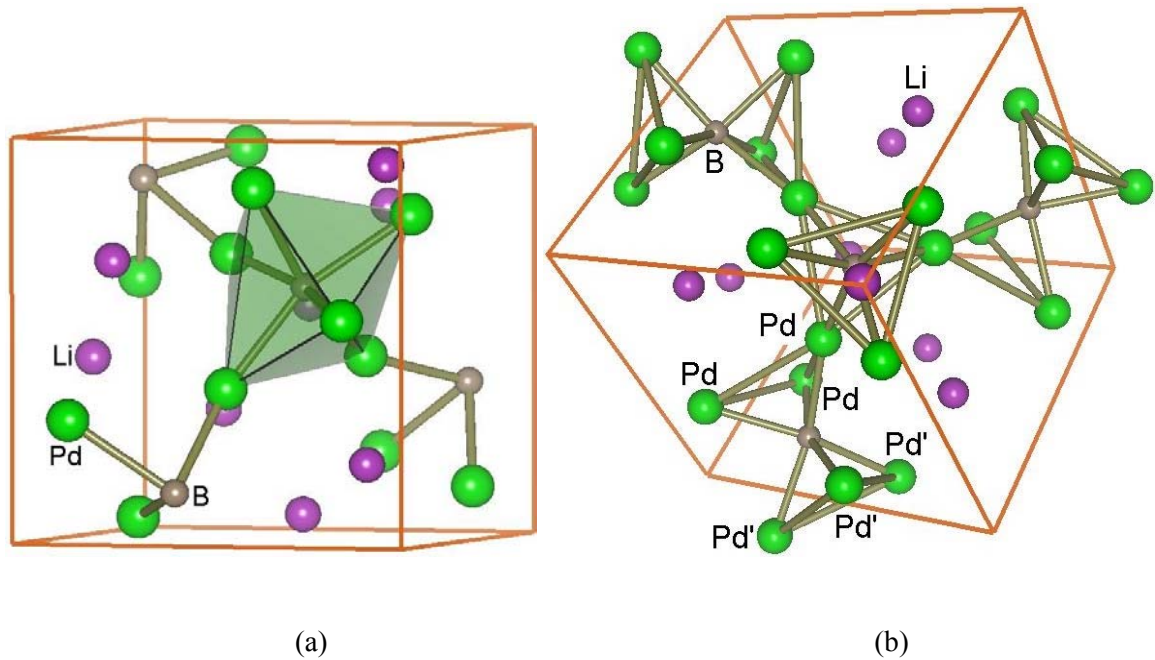

(a)  (b)

Fig.1: The crystal structure of $Li_2Pd_3B$ viewed in the (100) orientation (a) and in the (111) orientation (b). The distorted octahedron centered on the B atom shown in (a) can be viewed as comprising of two tetrahedra with the B atom shared between two neighboring tetrahedra and Pd atoms at the other three vertices of the tetrahedra. Even though the Pd atoms that form the two tetrahedra are equivalent, they have been marked as Pd and Pd' in (b) just to differentiate them.

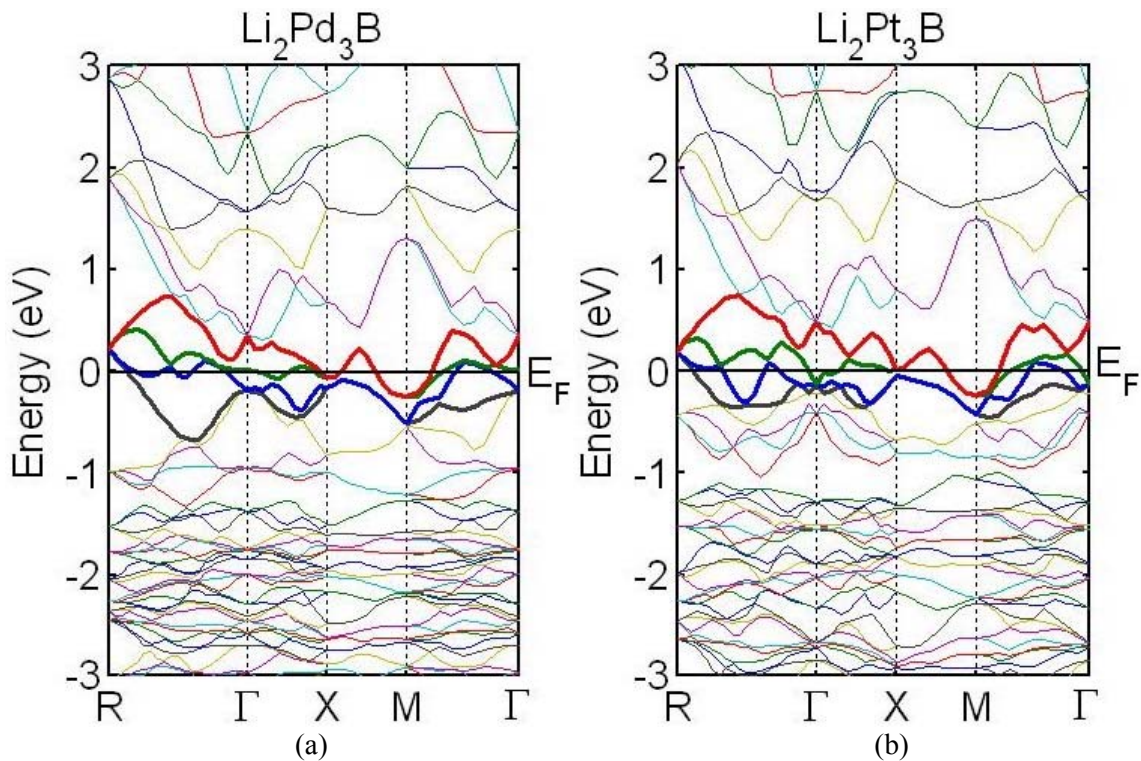

(a)  (b)

Fig.2: The band structure of $Li_2Pd_3B$ (a) and $Li_2Pt_3B$ (b). Four bands are seen to cross the Fermi level in both $Li_2Pd_3B$ and $Li_2Pt_3B$.

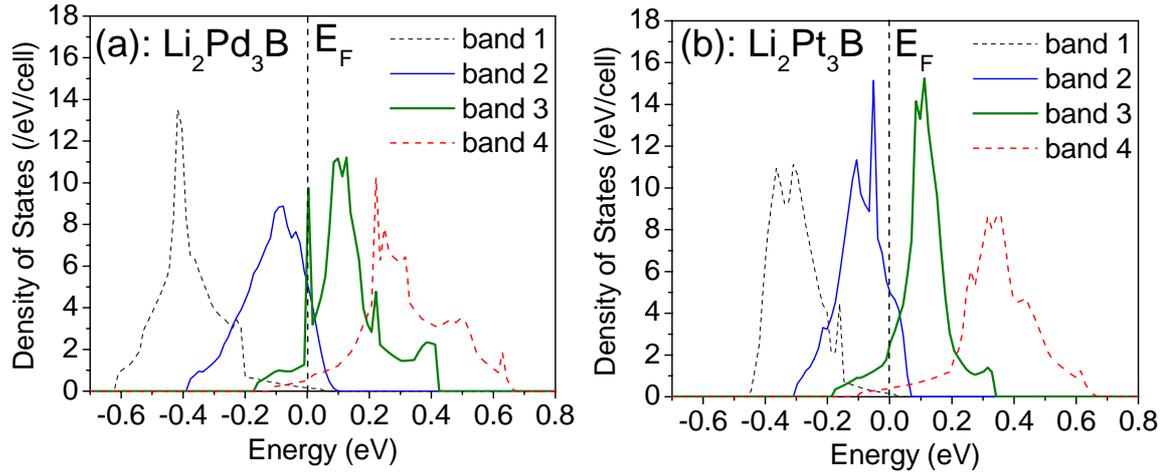

Fig.3: The partial density of states of the four bands crossing the Fermi level shown with thick lines in the band structure of $Li_2Pd_3B$ (a) and $Li_2Pt_3B$ (b). Only two bands out of the four are seen to contribute significantly to the partial DOS at the Fermi level.

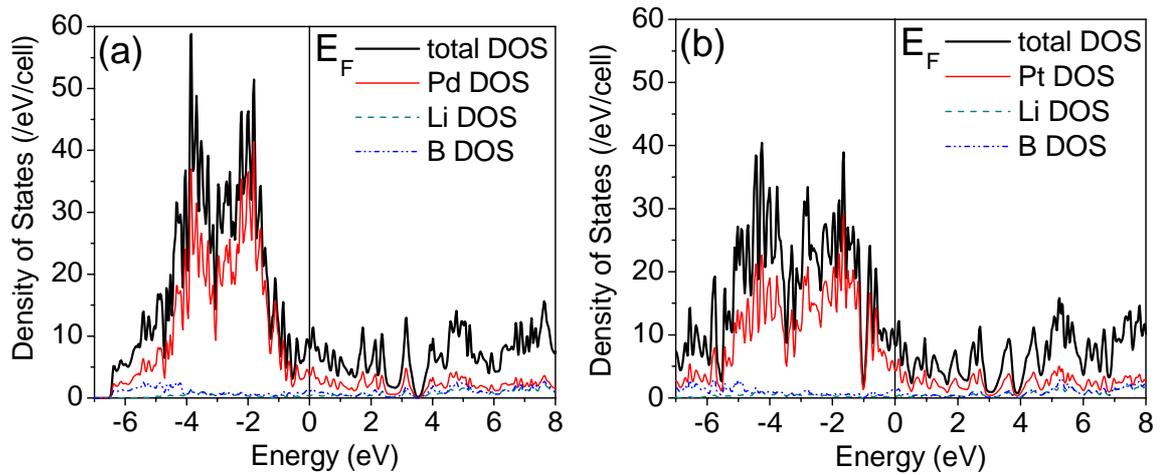

Fig.4: The total density of states of (a) $Li_2Pd_3B$ and (b) $Li_2Pt_3B$. Maximum contribution to the DOS comes from d orbitals of the Pd and Pt atoms. While B atoms contribute a small percentage, Li atom contribution is almost negligible. The maximum contribution to the total DOS at the Fermi level is from $d_{x^2-y^2}$ orbitals.

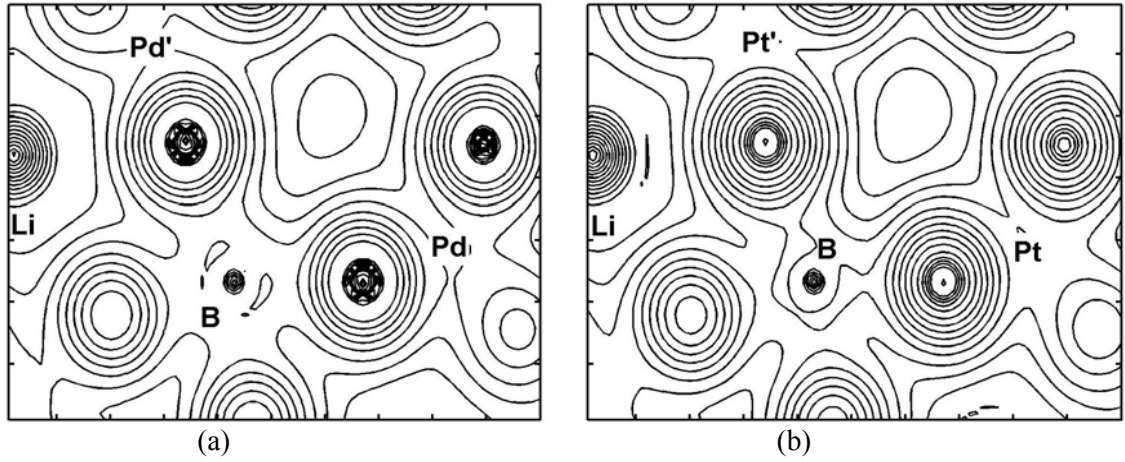

(a)          (b)

Fig.5: Electron density distribution in (a): $Li_2Pd_3B$ and (b): $Li_2Pt_3B$ compounds in the (311) plane. The Pd and Pt atoms are marked as Pd, Pd', Pt and Pt' following the convention of Fig.1. Li and B atoms that lie in the planes are also marked in the figure. Other Pd, Pt and Li atoms that are at some distance from the plane are also seen in the figure. It is seen that the magnitude of the electron density at the Pd, Pd' and Pt, Pt' atom sites is equal. Also, the electron density distribution is similar in both cases.

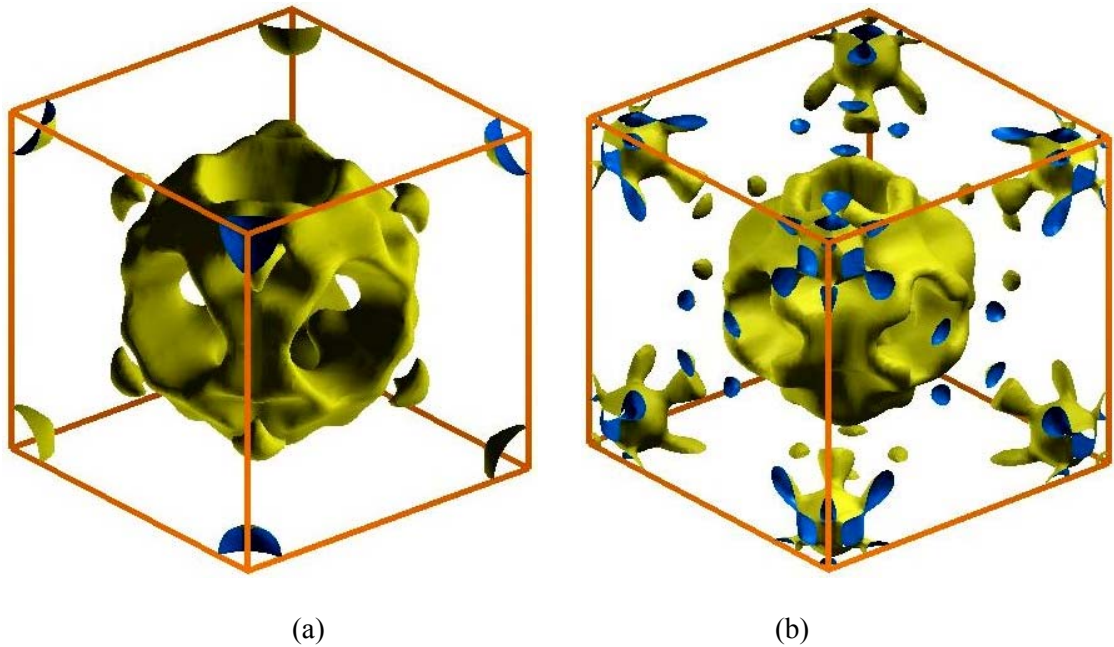

(a)          (b)

Fig.6: The Fermi surface corresponding to the second band that crosses the Fermi level in (a) $Li_2Pd_3B$ and (b) $Li_2Pt_3B$ compounds. Fermi surfaces have been plotted in the first Brillouin zone.

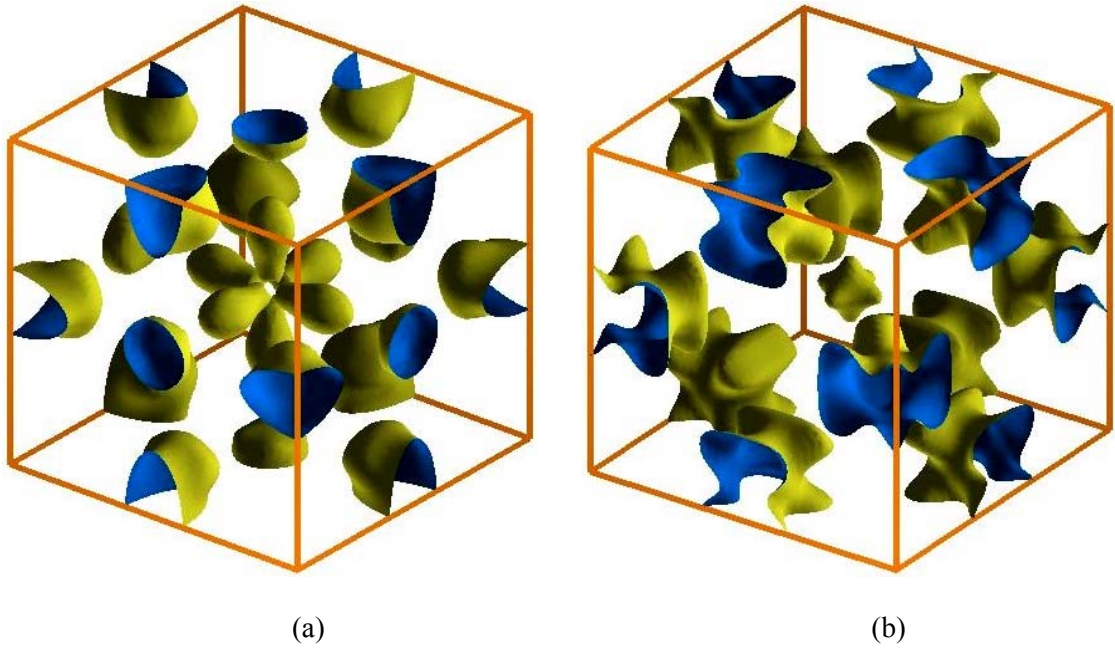

Fig.7: The Fermi surface corresponding to the third band that crosses the Fermi level in (a) $Li_2Pd_3B$ and (b) $Li_2Pt_3B$ compounds.

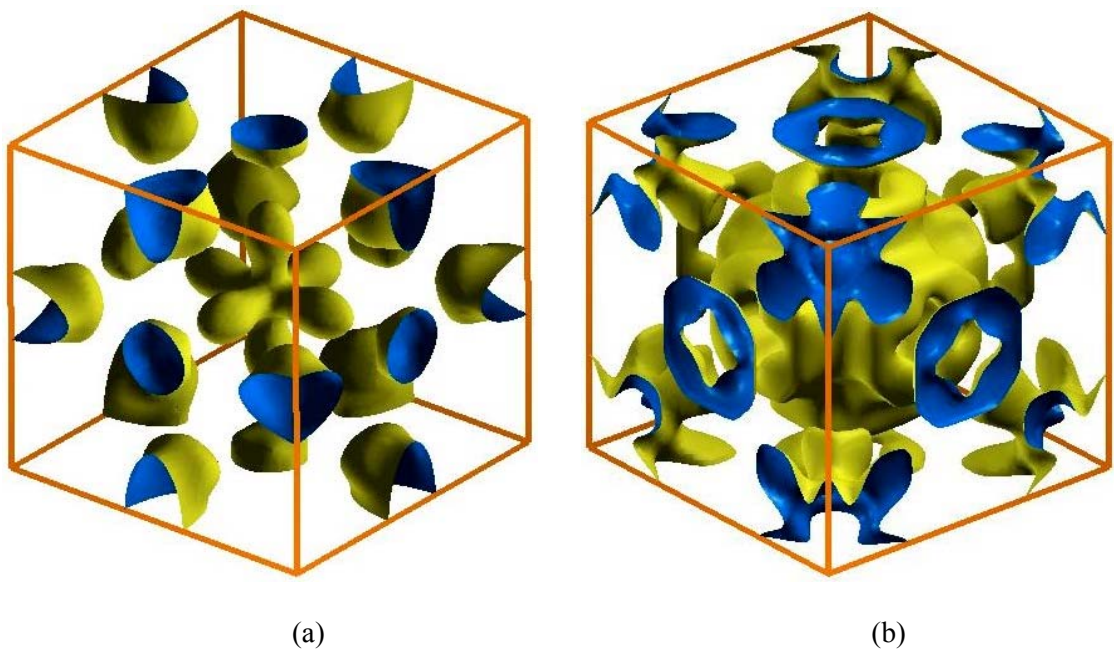

Fig.8: Topology of the constant energy surface at the energy of the van Hove singularity. (a) In $Li_2Pd_3B$: 4 meV above the Fermi energy in band 3 and (b) In $Li_2Pt_3B$: 52 meV below the Fermi energy in band 2.

18